\documentclass[12pt,a4paper]{revtex4}
\usepackage{graphicx}
\usepackage{amsmath}
\usepackage[usenames,dvipsnames]{color}

\newcommand {\threejot}[6]{\begin{pmatrix} #1\!\!&#2\!\!&#3\!\cr #4&#5&#6  \cr\end{pmatrix}}

\begin{document}

\title{Two center multipole expansion method: application to macromolecular systems}

\author{Ilia A. Solov'yov\footnote{On leave from the A.F. Ioffe Institute, St. Petersburg, Russia.
E-mail: ilia@fias.uni-frankfurt.de}, Alexander V. Yakubovich$^{\text
*}$, Andrey V. Solov'yov$^{\text *}$,
        and Walter Greiner}

\address{Frankfurt Institute for Advanced Studies,
Max von Laue Str. 1, 60438 Frankfurt am Main, Germany}

\begin{abstract}
We propose a new theoretical method for the calculation of the
interaction energy between macromolecular systems at large
distances. The method provides a linear scaling of the computing
time with the system size and is considered as an alternative to the
well known fast multipole method. Its efficiency, accuracy and
applicability to macromolecular systems is analyzed and discussed in
detail.
\end{abstract}



\maketitle

\section{Introduction}

In recent years, there has been {much} progress in simulating the
structure and dynamics of large molecules at the atomic level, which
may include up to thousands and millions of atoms
\cite{Schulten06,Schulten06_1,Schulten06_2,Meyer03}.  For example,
amorphous polymers may have segments each with {10000} atoms
\cite{Meyer03} which associate to form partially crystalline
lamellae, random coil regions, and interfaces between these regions,
each of which may contribute with special mechanical and chemical
properties to the system.

With increasing computer powers nowadays it became possible to study
molecular systems of enormous sizes which were not imaginable just
several years ago. For example in \cite{Schulten06} a molecular
dynamics simulations of the complete satellite tobacco mosaic virus
was performed which includes up to 1 million of atoms. In that paper
the stability of the whole virion and of the RNA core alone were
demonstrated, and a pronounced instability was shown for the capsid
without the RNA.


The study of structure and dynamics of macromolecules often implies
the calculation of the potential energy surface for the system. The
potential energy surface of a macromolecule carries a lot of useful
information about the system. For example from the potential energy
landscape it is possible to estimate the characteristic times for
the conformational changes \cite{YSSG06_EPJD,SYSG06_PRE,SYSG06_JETP}
and for fragmentation \cite{SYSG06_JETP_frag}. The potential
energy surface of a macromolecular system can be used for studying
the thermodynamical processes in the system such as phase
transitions \cite{SYSG06_JETP_ph_trans}. In proteins, the potential
energy surface is related to one of the most intriguing problems of
protein physics: protein folding
\cite{SYSG06_JETP_ph_trans,Chen05,Duan98,Liwo05,Ding02}.
The rate constants for complex biochemical reactions can also be
established from the analysis of the potential energy surface
\cite{ElsaPoster,ElsaPaper}.

The calculation of the potential energy surface and molecular
dynamics simulations often implies the evaluation of pairwise
interactions. The direct method for evaluating these potentials is
proportional to $\sim N^2$, where $N$ is the number of particles in
the system. This places a severe restraint on the treatable size of
the system. During the last two decades many different methods have
been suggested which provide a linear dependence of the
computational cost with respect to $N$
\cite{Gunsteren90,Petersen94,Ding92,White94,Schmidt91,Greengard85}.
The most widely used algorithm of this kind is the fast multipole
method (FMM)
\cite{Greengard85,Petersen94,Ding92,White94,Schmidt91,Boar2000,Pincus77,Gibbon02}.
The critical size of the system at which this method becomes
computationally faster than the exact method is accuracy dependent
and is very sensitive to the slope in the $N$ dependence of the
computational cost. In refs. \cite{Schmidt91,Ding92,Schulten92}
critical sizes ranging from $N\approx300$ to $N\approx30000$ have
been reported. Many discrepancies of the estimates in the critical
size arise from differences in the effort of optimizing the
algorithm and the underlying code. However, it is also important to
optimize the methods themselves with respect to the required
accuracy.

The FMM is based on the systematic organization of multipole
representations of a local charge distribution, so that each
particle interacts with local expansions of the potential.
Originally FMM was introduced in \cite{Greengard85} by Greengard and
Rokhlin. Later, Greengard's method has been implemented in various
forms. Schmidt and Lee \cite{Schmidt91} have produced a version
based upon the spherical multipoles for both periodic and
nonperiodic systems. Zhou and Johnson have implemented the FMM for
use on parallel computers \cite{Zhao91}, while Board {\it et al}
have reported both serial and parallel versions of the FMM
\cite{Schulten92}.

Ding {\it et al} introduced a version of the FMM that relies upon
Cartesian rather than spherical multipoles \cite{Ding92}, which they
applied to very large scale molecular dynamics calculations.
Additionally they modified Greengard's definition of the nearest
neighbors to increase the proportion of interactions evaluated via
local expansions. Shimada {\it et al} also developed a cartesian
based FMM program \cite{Shimada94}, primarily to treat periodic
systems described by molecular mechanics potentials. In both cases
only low order multipoles were employed, since high accuracy was not
sought.

In the present paper we suggest a new method for calculating the
interaction energy between macromolecules. Our method also provides
a linear scaling of the computational costs with the size of the
system and is based on the multipole expansion of the potential.
However, the underlying ideas are quite different from the FMM.

Assuming that atoms from different macromolecules interact via a
pairwise Coulomb potential, we expand the potential around the
centers of the molecules and build a two center multipole expansion
using bipolar harmonics algebra. Finally, we obtain a general
expression which can be used for calculating the energy and forces
between the fragments. This approach is different from the one used
in the FMM, where the so-called translational operators were used to
expand the potential around a shifted center. Note that the final
expression, which we suggest in our theory was not discussed before
within the FMM. Similar expressions were discussed since the earlier
50's (see e.g. \cite{Buehler51,Joslin84,Gupta81,Wolf68}). In these
papers the two center multipole expansion was considered as a new
form of Coulomb potential expansion, but the expansion was never
{applied to} the study of macromolecular systems.

We consider the interaction of macromolecules via Coulomb potential
since this is the only long-range interaction in macromolecules,
which is important for the description of the potential energy
surface at large distances. Other interaction terms in
macromolecular systems are of the short-range type and become
important when macromolecules get close to each other
\cite{SYSG06_JETP_frag}. At large distances these terms can be
neglected.

In the present paper we show that the method based on the two center
multipole expansion can be used for computing the interaction energy
between complex macromolecular systems. In section \ref{theory} we
present the formalism which lies behind the two center multipole
expansion method. In subsection \ref{compCoulomb} we analyze the
behavior of the computation cost of this method and establish the
critical sizes of the system, when the two center multipole
expansion method demands less computer time than the exact energy
calculation approach. {In subsection \ref{compFMM} we compare the
results of our calculation with the results obtained within the
framework of the FMM}. In section \ref{accur} we discuss the
accuracy of the two center multipole expansion method.

\section{Two center multipole expansion method}
\label{theory}

In this section we present the formalism, which underlies the two
center multipole expansion method, which will be further referred to
as the TCM method.

Let us consider two multi atomic systems, which we will denote as
$A$ and $B$. The pairwise Coulomb interaction energy of those
systems can be written as follows:

\begin{equation}
U=\sum_{i=1}^{N_A}\sum_{j=1}^{N_B}\frac{q_iq_j}{\left|{\bf
R_j^B}-{\bf
R_i^A}\right|}=\sum_{i=1}^{N_A}\sum_{j=1}^{N_B}\frac{q_iq_j}{\left|{\bf
R_0}+{\bf r_j^B}-{\bf r_i^A}\right|}, \label{Coulomb}
\end{equation}

\noindent where $N_A$ and $N_B$ are the total number of atoms in
systems $A$ and $B$ respectively, $q_i$ and $q_j$ are the charges of
atoms $i$ and $j$ from the system $A$ and $B$ respectively, ${\bf
R_0}$ is the vector interconnecting the center of system $A$ with
the center of system $B$, ${\bf r_i^A}$ and ${\bf r_j^B}$ are the
vectors describing the position of charges $i$ and $j$ with respect
to the centers of the system $A$ and $B$ respectively. The centers
of both systems can be any suitable points of each of the molecules.
It is natural to define them as the centers of mass of the
corresponding systems, but in some cases another choice might be
more convenient (see for example \cite{SYSG06_JETP_frag}, where we
have applied the TCM method for studying fragmentation of alanine
dipeptide).

Expression (\ref{Coulomb}) can be expanded into a series of
spherical harmonics. The expansion depends on the vectors ${\bf
R_0}$, ${\bf r_i^A}$ and ${\bf r_j^B}$. In the present paper we
consider the case when

\begin{equation}
\left|{\bf R_0}\right|>\left|{\bf r_j^B}\right|+\left|{\bf
r_i^A}\right| \label{mainCond}
\end{equation}

\noindent holds for all $i$ and $j$. This particular case is
important, because it describes well separated charge distributions,
and can be used for modeling the interaction between complex objects
at large distances. In this case the expansion of (\ref{Coulomb})
reads as \cite{Varshalovich}:

\begin{eqnarray}
\label{SpherHarm_exp1}
\nonumber
\frac{q_iq_j}{\left|{\bf R_0}+{\bf r_j^B}-{\bf
r_i^A}\right|}&=&q_iq_j\sum_{L=0}^{\infty}\sum_{M=-L}^{L}\frac{4\pi
}{2L+1}\frac{\left|{\bf r_j^B}-{\bf r_i^A}\right|^L}{R_0^{L+1}}\\
&&Y_{LM}^{*}\left(\Theta_{{\bf r_j^B}-{\bf r_i^A}},\Phi_{{\bf
r_j^B}-{\bf r_i^A}}\right) Y_{LM}\left(\Theta_{{\bf R_0}},\Phi_{{\bf
R_0}}\right)
\end{eqnarray}

According to \cite{Varshalovich} the function

$$
\left|{\bf r_j^B}-{\bf r_i^A}\right|^{L}Y_{LM}^{*}\left(\Theta_{{\bf
r_j^B}-{\bf r_i^A}}, \Phi_{{\bf r_j^B}-{\bf r_i^A}}\right)
$$

\noindent can be expanded into series of bipolar harmonics:

\begin{eqnarray}
\label{caseIII} \nonumber &&\left|{\bf r_j^B}-{\bf
r_i^A}\right|^{L}Y_{LM}^{*}\left(\Theta_{{\bf r_j^B}-{\bf r_i^A}},
\Phi_{{\bf r_j^B}-{\bf
r_i^A}}\right)=\sqrt{4\pi(2L+1)!}\sum_{{l_1,l_2=0}\atop{l_1+l_2=L}}^{L}
\\
&&(-1)^{l_2}\frac{\left(r_i^A\right)^{l_1}\left(r_j^B\right)^{l_2}}{\sqrt{(2l_1+1)!(2l_2+1)!}}\left(Y_{l_1}(\Theta_{r_i^A},\Phi_{r_i^A})\otimes
Y_{l_2}(\Theta_{r_j^B},\Phi_{r_j^B})\right)_{LM}^{*}
\end{eqnarray}

\noindent where a bipolar harmonic is defined as follows:

\begin{eqnarray}
\label{bipolar}
\nonumber
\left(Y_{l_1}(\Theta_{r_1},\Phi_{r_1})\right.&\otimes&
\left.Y_{l_2}(\Theta_{r_2},\Phi_{r_2})\right)_{LM}=\\
&&\sum_{m_1,m_2}C_{l_1m_1l_2m_2}^{LM}Y_{l_1m_1}(\Theta_{r_1},\Phi_{r_1})Y_{l_2m_2}(\Theta_{r_2},\Phi_{r_2}).
\end{eqnarray}

\noindent Here $C_{l_1m_1l_2m_2}^{LM}$ are the Clebsch-Gordan
coefficients, which can be transformed to the
$3j$- symbol notation as follows:

\begin{equation}
\label{3j}
C_{l_1m_1l_2m_2}^{LM} = (-1)^{l_1-l_2+M}\sqrt{2L+1}
\threejot{l_1}{l_2}{L}{m_1}{m_2}{-M}
\end{equation}

Using equations (\ref{caseIII}),(\ref{3j}) and (\ref{bipolar}) we
can rewrite expansion (\ref{SpherHarm_exp1}) as follows:

\begin{eqnarray}
\label{SpherHarm_exp2}
\nonumber
\frac{q_iq_j}{\left|{\bf R_0}+{\bf r_j^B}-{\bf
r_i^A}\right|}&=&q_iq_j\sum_{{l_1,l_2=0}\atop{l_1+l_2=L}}^{\infty}\sum_{{m_1=-l_1}\atop{m_2=-l_2}}^{l_1,l_2}
(-1)^{l_1+M}
\sqrt{\frac{(4\pi)^3(2L)!}{(2l_1+1)!(2l_2+1)!}}
\threejot{l_1}{l_2}{L}{m_1}{m_2}{-M}\\
&&
\frac{\left(r_i^A\right)^{l_1}\left(r_j^B\right)^{l_2}}{R_0^{L+1}}
Y_{l_1m_1}(\Theta_{r_i^A},\Phi_{r_i^A})Y_{l_2m_2}(\Theta_{r_j^B},\Phi_{r_j^B})
Y_{LM}^{*}\left(\Theta_{R_0},\Phi_{R_0}\right)
\end{eqnarray}

The multipole moments of systems $A$ and $B$ are defined as follows:

\begin{eqnarray}
\label{multipole}
Q_{l_1m_1}^{A}&=&\sum_{i=1}^{N_A}q_i\left(r_i^{A}\right)^{l_1}\sqrt{\frac{4\pi}{2l_1+1}}Y_{l_1m_1}(\Theta_{r_i^{A}},\Phi_{r_i^{A}})\\
\nonumber
Q_{l_2m_2}^{B}&=&\sum_{j=1}^{N_B}q_j\left(r_j^{B}\right)^{l_2}\sqrt{\frac{4\pi}{2l_2+1}}Y_{l_2m_2}(\Theta_{r_j^{B}},\Phi_{r_j^{B}}).
\end{eqnarray}

Summing equation (\ref{SpherHarm_exp2}) over $i$ and $j$, and
accounting only for the first $L_{max}$ multipoles in both systems,
we obtain:

\begin{eqnarray}
\label{Formula}
\nonumber
U_{mult}&=&\sum_{{l_1,l_2=0}\atop{l_1+l_2=L}}^{L_{max}}\sum_{{m_1=-l_1}\atop{m_2=-l_2}}^{l_1,l_2}
\frac{(-1)^{l_1+M}}{R_0^{L+1}}
\sqrt{\frac{4\pi(2L)!}{(2l_1)!(2l_2)!}}
\threejot{l_1}{l_2}{L}{m_1}{m_2}{-M}\\
&&
Q_{l_1m_1}^{A}Q_{l_2m_2}^{B}Y_{LM}^{*}\left(\Theta_{R_0},\Phi_{R_0}\right).
\end{eqnarray}

This expression describes the electrostatic energy of the system in
terms of a two center multipole expansion. Note, that this expansion
is only valid when the condition $R_0>r_i^A+r_j^B$ holds for all $i$
and $j$, otherwise more sophisticated expansions have to be
considered, which is beyond the scope of the present paper.

Summation in equation (\ref{Formula}) is performed over
$l_1,l_2\in[0...L_{max}]$; $m_1\in[-l_1...l_1]$;
$m_2\in[-l_2...l_2]$, and the condition $M=m_1+m_2$ holds. $L_{max}$
is the principal multipole number, which determines the number of
multipoles in the expansion.

\section{Computational efficiency}
\label{compeff}

\subsection{Comparison with direct Coulomb interaction method}
\label{compCoulomb}

In this section we discuss the computational efficiency of
the TCM method. For this purpose we have analyzed the time required
for computing the Coulomb interaction energy between two systems of
charges and the time required for the energy calculation within the
framework of the TCM method for different system sizes, and for
different values of the principal multipole number.

For the study of the computational efficiency of the TCM method we
{have} considered the interaction between two systems (we denote
them as $A$ and $B$) of randomly distributed charges, for which the
condition eq. (\ref{mainCond}) holds. The charges in both systems
were randomly distributed within the spheres of radii $R_A=1.0\cdot
N_A^{1/3}$ and $R_B=1.0\cdot N_B^{1/3}$ respectively and the
distance between the centers of mass of the two systems was chosen
as $R_0=3/2(R_A+R_B)$.

The computational time needed for the energy calculations is
proportional to the number of operations required. Thus, the time
needed for the Coulomb energy calculation (CE calculation) can be
estimated as:

\begin{equation}
\label{timeCoul}
\tau_{Coul}=\alpha_{Coul}N_A N_B\sim N^2
\end{equation}

\noindent where $\alpha_{Coul}$ is a constant depending on the
computer processor power and on the efficiency of the code, $N_A\sim
N_B\sim N$. From equation (\ref{timeCoul}) it follows that the
computational cost of the CE calculation grows proportionally to the
second power of the system size.

For large systems the TCM method becomes more efficient because it
provides a linear scaling with the system size. The time needed for
the energy calculation reads as follows:


\begin{eqnarray}
\label{timeMult} \tau_{mult}(L_{max})&=&\beta
N+\sum_{l_1=0}^{L_{max}}\sum_{l_2=0}^{L_{max}}\sum_{m_1=-l_1}^{l_1}\sum_{m_2=-l_2}^{l_2}\left(N_A\tau_{l_1,m_1}+N_B\tau_{l_2,m_2}\right)\approx\\
\nonumber &\approx&\alpha_{mult}L_{max}(1+L_{max})^3N,
\end{eqnarray}

\noindent where the first term, $\beta N$, corresponds to the
computer time needed for allocating arrays in memory and tabulating
the computationally expensive functions like $\cos(\Phi)$ and
$\exp(im\Phi)$. $\tau_{l,m}$ is the time needed for evaluation of
the spherical harmonic at given $l$ and $m$, and $\alpha_{mult}$ is
a numerical coefficient, which depends on the processor power and on
the efficiency of the code. In general it is different from
$\alpha_{Coul}$.

\begin{figure}[h]
\includegraphics[scale=0.8,clip]{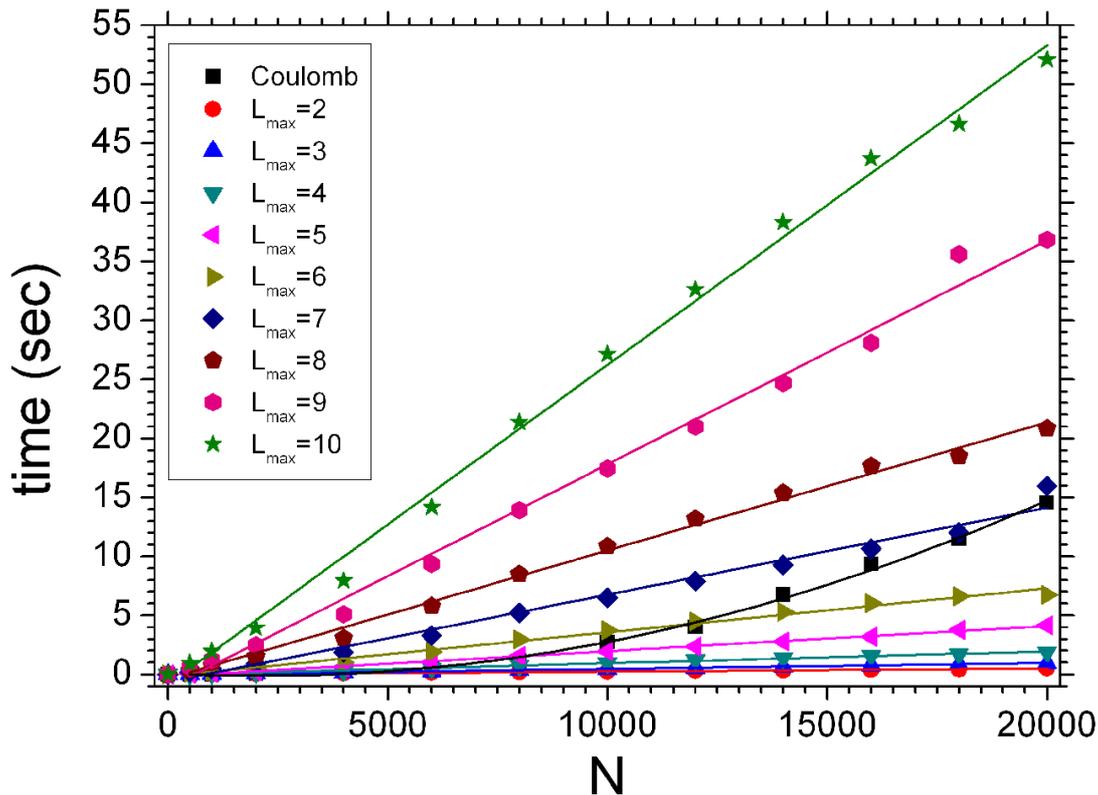}
\caption{Time needed for energy calculation as a function of the
system size.} \label{efficiency}
\end{figure}

In Fig.~\ref{efficiency} we present the dependencies of the computer
time needed for the CE calculation (squares) and for the computation
of energy within the TCM method for different values of the
principal multipole number as a function of system size. This data
was obtained on a 1.8 GHz 64-bit AMD Opteron-244 computer.

From Fig.~\ref{efficiency} it is clear that the time needed for the
CE calculation has a prominent parabolic trend that is consistent
with the analytical expression (\ref{timeCoul}). The fitting
expression which describes this dependance is given in the first row
of Tab.~\ref{peresech}. At large $N$ the $N^2$ term becomes dominant
and the other two terms can be neglected. Thus,
$\alpha_{Coul}\approx4.46\cdot10^{-8}$ $(sec)$.

\begingroup
\begin{table*}[h]
\caption{Fitting expressions for the computational time needed for
the CE calculation and for the energy computation within the TCM
method at different values of the principal multipole number,
$L_{max}$ (second column). System sizes, for which the Coulomb
energy calculation becomes more computer time demanding at a given
value of $L_{max}$ are shown in the third column.} \label{peresech}

\begin{ruledtabular}
\begin{tabular}{ccc}

     \multicolumn{1}{c}{$L_{max}$} &
     \multicolumn{1}{c}{$\tau(N)$ (sec.)} &
     \multicolumn{1}{c}{$N_{max}$}\\

\hline

Coulomb & $0.11736-0.0002N+4.6768\cdot10^{-8}N^2$ & - \\
2  & $-0.01986+3.0\cdot10^{-5}N$  & 4223 \\
3  & $-0.03159+5.0\cdot10^{-5}N$  & 4662 \\
4  & $-0.04714+1.0\cdot10^{-4}N$  & 5809 \\
5  & $-0.16054+2.1\cdot10^{-4}N$  & 8026 \\
6  & $-0.14710+3.7\cdot10^{-4}N$  & 11704 \\
7  & $-0.59675+7.4\cdot10^{-4}N$  & 19308 \\
8  & $-0.35383+10.9\cdot10^{-4}N$ & 27212 \\
9  & $-1.15856+1.9\cdot10^{-3}N$  & 44286 \\
10 & $-0.83688+2.71\cdot10^{-3}N$ & 61892
\end{tabular}
\end{ruledtabular}
\end{table*}
\endgroup

The fitting expressions which describe the time needed for the
energy computations within the TCM method at different values of the
principal multipole number are given in Tab.~\ref{peresech}, rows
2-10. These expressions were obtained by fitting the data shown in
Fig.~\ref{efficiency}. Note the linear dependence on $N$. The
numerical coefficient in all expressions correspond to the factor
$\alpha_{mult}L_{max}(1+L_{max})^3$ in equation (\ref{timeMult}).
The fitting expressions in Tab.~\ref{peresech} were
obtained by fitting of data obtained for systems with large number
of particles (see Fig.~\ref{efficiency}). Therefore these
expressions are applicable when $N\gg1$.

From equations presented in Tab.~\ref{peresech} it is possible to
determine the critical system sizes at which the TCM method becomes
less computer time demanding then the CE calculation. The critical
system sizes calculated for different principal multipole numbers
are shown in the third column of Tab.~\ref{peresech}. These sizes
correspond to the intersection points of the parabola describing the
time needed for the CE calculation with the straight lines describing
the computational time needed for the TCM method. In
Fig.~\ref{efficiency} one can see six intersection points for
$L_{max}=2-7$.

\begin{figure}[h]
\includegraphics[scale=0.8,clip]{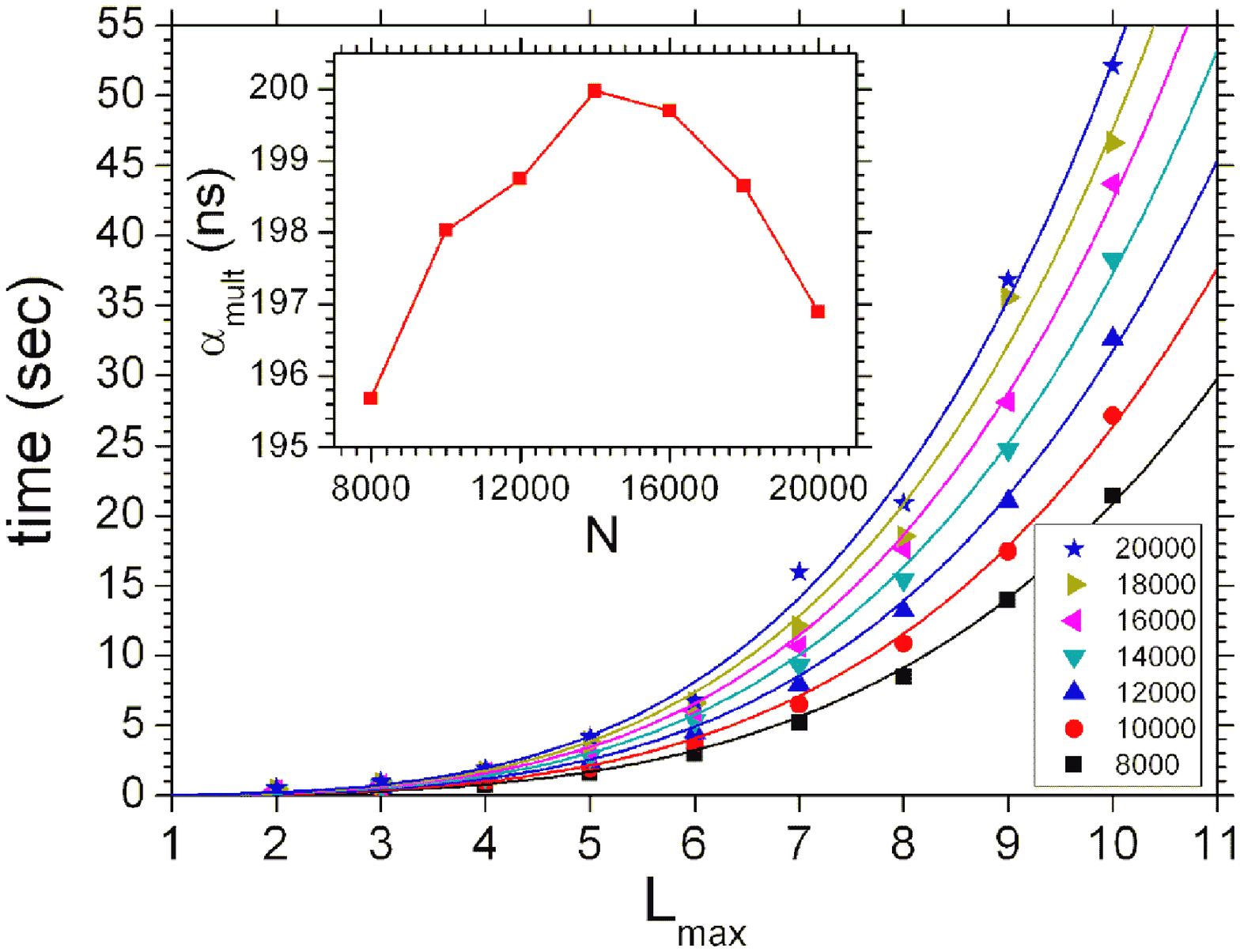}
\caption{Time needed for the calculation of energy of the systems of
different sizes computed within the framework of the TCM method as a
function of the principal multipole number $L_{max}$. In the inset
we plot $\alpha_{mult}$ as a function of the system size.}
\label{Ldep}
\end{figure}

From equation (\ref{timeMult}), it follows that computation time of
the energy within the framework of the TCM method grows as the power
of 4 with increasing $L_{max}$. To stress this fact, in
Fig.~\ref{Ldep} we present the dependencies of the computation time
obtained within the TCM method at different system sizes as a
function of principal multipole number. All curves shown in
Fig.~\ref{Ldep} can be perfectly fitted by the analytical expression
(\ref{timeMult}). In the inset to Fig.~\ref{Ldep}, we plot the
dependence of the fitting coefficient $\alpha_{mult}$ as a function
of the system size. From this plot it is seen that $\alpha_{mult}$
varies only slightly for all system sizes considered, being equal to
$(1.982\pm0.015)\cdot10^{-7}$ $(sec)$.

Thus, the expression for the time needed for the energy calculation
within the framework of the TCM method reads as:

\begin{equation}
\label{timeMult_fin}
\tau_{mult}(L_{max})\approx1.98\cdot10^{-7}L_{max}(1+L_{max})^3N.
\end{equation}

\noindent Note, that $\alpha_{mult}=1.98\cdot10^{-7}$ $(sec)$ is
larger than $\alpha_{Coul}\approx4.46\cdot10^{-8}$ $(sec)$, since in
one turn of the TCM method more algebraic operations have to be
done, than in one turn of the CE calculation.

From the analysis performed in this section it is clear that the TCM
method can give a significant gain in the computation time. However,
at larger principal multipole numbers ($L_{max}=8,9,10$) this method
can compete with the CE calculation only at system sizes greater
than 27000-61000 atoms. The accounting for higher multipoles is
necessary if the distance between two interacting systems becomes
comparable to the size of the systems. In the next section we
discuss in detail the accuracy of the TCM method and identify
situations in which higher multipoles should be accounted for.

{\subsection{Comparison with the fast multipole method}
\label{compFMM}

\begin{figure}[h]
\includegraphics[scale=0.8,clip]{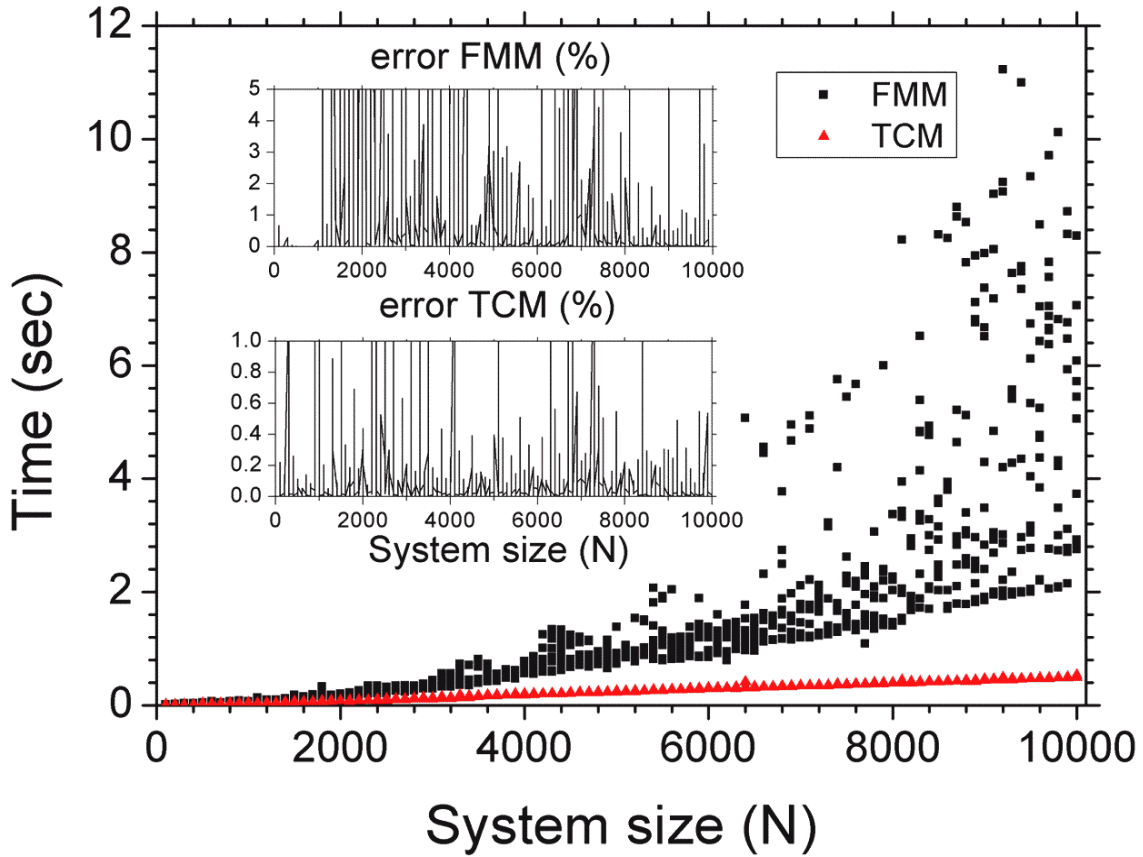}
\caption{{Time needed for the calculation of the interaction energy
between two systems as a function the total number of particles
calculated within the framework of the TCM method (triangles) and
within the framework of the FMM (squares). In the upper and lower
insets we plot the relative error of the FMM and of the TCM methods
as a function of the system size respectively.}} \label{FMM_vs_TCM}
\end{figure}

The fast multipole method (FMM) \cite{Greengard85,Boar2000,Pincus77}
is a well known method for calculating the electrostatic energy in a
multiparticle system, which provides a linear scaling of the
computing time with the system size. In order to stress the
computational efficiency of the TCM method in this section we
compare the time required for the energy calculation within the
framework of the FMM and using the TCM method.

To perform such a comparison we used an adaptive FMM library, which
has been implemented for the Coulomb potential in three dimensions
\cite{Gibbon02,FMMcode}. We have generated two random charge
distributions of different size and calculated the interaction
energies between them as well as the required computation time using
the FMM and the TCM methods. As in the previous section the charges
in both systems were randomly distributed within the spheres of
radii $R_A=1.0\cdot N_A^{1/3}$ and $R_B=1.0\cdot N_B^{1/3}$
respectively and the distance between the center of mass of the two
systems was chosen as $R_0=3/2(R_A+R_B)$.

In Fig.~\ref{FMM_vs_TCM} we present the comparison of the computer
time needed for the FMM calculation (squares) and for the
computation of energy within the TCM method (triangles) as a
function of system size. These data were obtained on an Intel(R)
Xeon(TM) CPU 2.40GHz computer. In the upper and lower insets of
Fig.~\ref{FMM_vs_TCM} we show the relative error of the FMM and of
the TCM methods as a function of the system size respectively, which
is defined as follows:

\begin{equation}
\eta_{method}=\frac{|U_{coul}-U_{method}|}{|U_{coul}|}\cdot100\%.
\label{eq:error_FMM_TCM}
\end{equation}

\noindent Here $method$ indicates the FMM or the TCM methods. For
comparing the efficiency of the two methods we have considered
different charge distributions within the size range of 100 to 10000
particles. Each point in Fig.~\ref{FMM_vs_TCM} corresponds to a
particular charge distribution. For each system size ten different
charge distributions were used. The time of the FMM calculation
depends on the charge distribution, as is clearly seen in
Fig.~\ref{FMM_vs_TCM}. Note that for a given system size the
calculation time of the FMM can change by more than a factor of 5,
depending on the charge distribution (see points for $N=10000$ in
Fig.~\ref{FMM_vs_TCM}).

For all system sizes FMM requires some minimal computer time for
calculating the energy of the system, which increases with the
growth of system size (see Fig.~\ref{FMM_vs_TCM}). The comparison of
the minimal FMM computation time with the computation time required
for the TCM method shows that the TCM method appears to be
significantly faster than the FMM. For $N=10000$ FMM requires at
least 2.15 seconds to compute the energy, while TCM method requires
0.53 seconds, being approximately 4 times faster.

The results of the TCM method calculation shown in
Fig.~\ref{FMM_vs_TCM}, were obtained for $L_{max}=2$. The analysis
of relative errors presented in the inset to Fig.~\ref{FMM_vs_TCM}
shows that with this principal multipole number it is possible to
calculate the energy between two systems with an error of less than
1 \% for almost arbitrary charge distribution. Note that for the
same charge distributions the error of the FMM is much more, being
about 5 \% in almost all of the considered systems. This allows us
to conclude that the TCM method is more efficient and more accurate
than the classical FMM.

It is important to mention that in the traditional implementation,
FMM calculates the total electrostatic energy of the system while
TCM method was developed for studying interaction energy between
system fragments. It is possible to modify the FMM to study only
interaction energies between different parts of the system.
However, the computation cost of the modified FMM is expected to be
higher than of the TCM method. This happens because, within the
framework of the modified FMM method, the field created by one
fragment of the system should be expanded in the multipole series
and the interactions of the resulting multipole moments with the
charges from another fragment should be calculated. Thus the
computation cost of this method will be proportional to $N_A\cdot
N_B$, where $N_A$ and $N_B$ are the number of particles in two
fragments, while the TCM method is proportional to $N_A+N_B$. The
computation cost of the modified version of the FMM depends
quadratically on the size of the system, because in this method the
interacting fragments should be considered as two independent cells,
while traditional FMM uses a hierarchical subdivision of the whole
system into cells to achieve linear scaling.

So far we have considered only the interaction between two multi
particle systems in {\it vacuo}, and demonstrated the efficiency of
the TCM method in this case, although the TCM method can also be
applied to the larger number of interacting systems. The study of
structure and dynamics of biomolecular systems consisting of several
components (i.e an ensemble of proteins, DNA, macromolecules in
solution) is a separate topic, which is beyond the scope of this
paper and deserves a separate investigation.
}

\section{Accuracy of the TCM method. Potential energy surface for porcine pancreatic trypsin/soybean trypsin inhibitor
complex.} \label{accur}

We have calculated the interaction energy between two proteins
within the framework of the TCM method and compared it with the
exact Coulomb energy value. On the basis of this comparison we
{have} concluded about the accuracy of the TCM method.

In the present paper we {have} studied the interaction energy
between the porcine pancreatic trypsin and the soybean trypsin
inhibitor proteins (Protein Data Bank (PDB) \cite{Protbase} entry
1AVW \cite{Song98}). Trypsins are digestive enzymes produced in the
pancreas in the form of inactive trypsinogens. They are then
secreted into the small intestine, where they are activated by
another enzyme into trypsins. The resulting trypsins themselves
activate more trypsinogens (autocatalysis). Members of the trypsin
family cleave proteins at the carboxyl side (or "C-terminus") of the
amino acids lysine and arginine. Porcine pancreatic trypsin is a
archetypal example. Its natural non-covalent inhibitor (porcine
pancreatic trypsin inhibitor) inhibits the enzyme's activity in the
pancreas, protecting it from self-digestion.

Trypsin is also inhibited non-covalently by the soybean trypsin
inhibitor from the soya bean plant, although this inhibitor is
unrelated to the porcine pancreatic trypsin inhibitor family of
inhibitors. Although the biological function of the soybean trypsin inhibitor is
mostly unknown it is assumed to help defend the plant from
insect attack by interfering with the insect digestive system.

\begin{figure}[h]
\includegraphics[scale=0.8,clip]{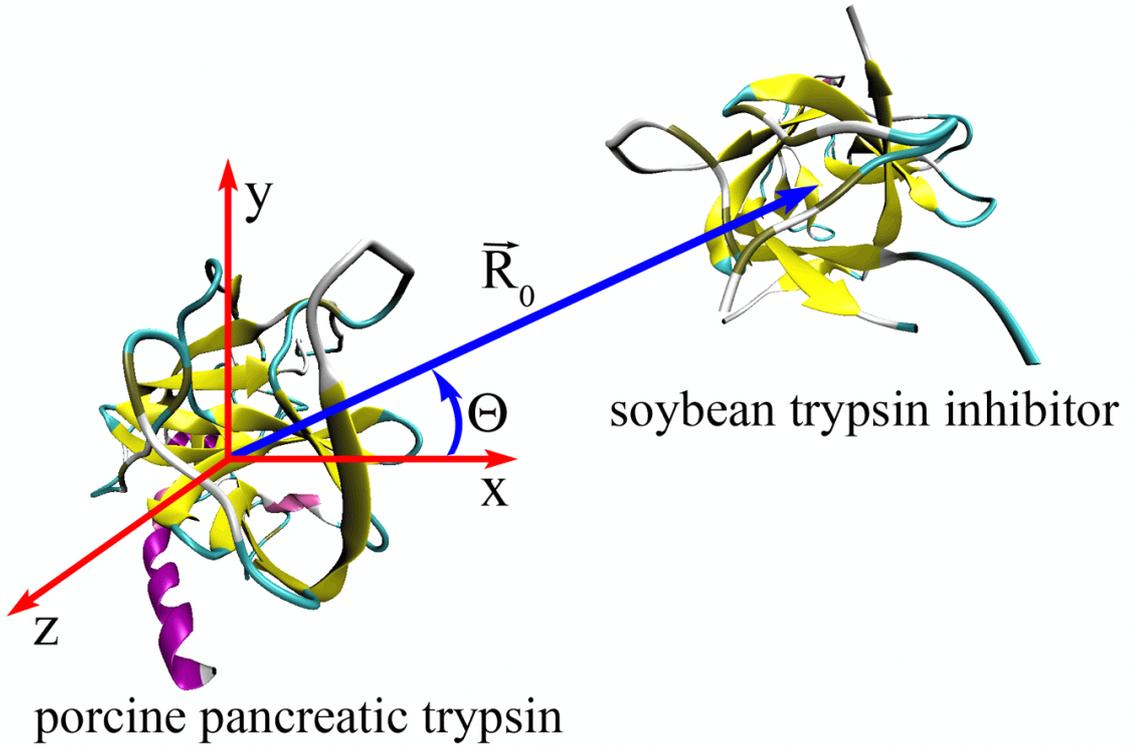}
\caption{Structure of the porcine pancreatic trypsin and soybean
trypsin inhibitor with the coordinate frame used for the energy
computation. {Figure has been rendered with help of the VMD
visualization package \cite{Humphrey96}}} \label{belki}
\end{figure}

The structure of both proteins is shown in Fig.~\ref{belki}. The
coordinate frame used for our computations is marked in the figure.
This coordinate frame is consistent with the standard coordinate
frame used in the PDB.

We use this particular example as a model system in order to
demonstrate the possible use of the TCM method. Therefore environmental
effects are omitted and we consider
only the protein-protein interaction {\it in vacuo}. The porcine
pancreatic trypsin and the soybean trypsin inhibitor include 223 and
177 amino acids respectively. Both proteins include 5847 atoms. Thus
for such system size the TCM method is faster than the CE
calculation if $L_{max}\le4$ (see Tab.~\ref{peresech}).

We have calculated the interaction energy between the porcine
pancreatic trypsin and soybean trypsin inhibitor as a function of
distance between the centers of masses of the fragments, ${\vec
R_{0}}$, and the angle $\Theta$, which is determined as the angle
between the x-axis and the vector ${\vec R_{0}}$ (see
Fig.~\ref{belki}). We have assumed that the porcine pancreatic trypsin is fixed
in space at the center of the coordinate frame and have restricted
${\vec R_{0}}$ to the (xy)-plane. Of course, the two degrees of
freedom considered are not sufficient for a complete description of
the mutual interaction between the two systems. For this purpose at
least six degrees of freedom are needed. However for our example of
the energy calculation of the porcine pancreatic trypsin/soybean
trypsin inhibitor complex within the framework of the TCM method the
two coordinates ${\vec R_{0}}$ and $\Theta$ are sufficient.

\begin{figure}[h]
\includegraphics[scale=0.7,clip]{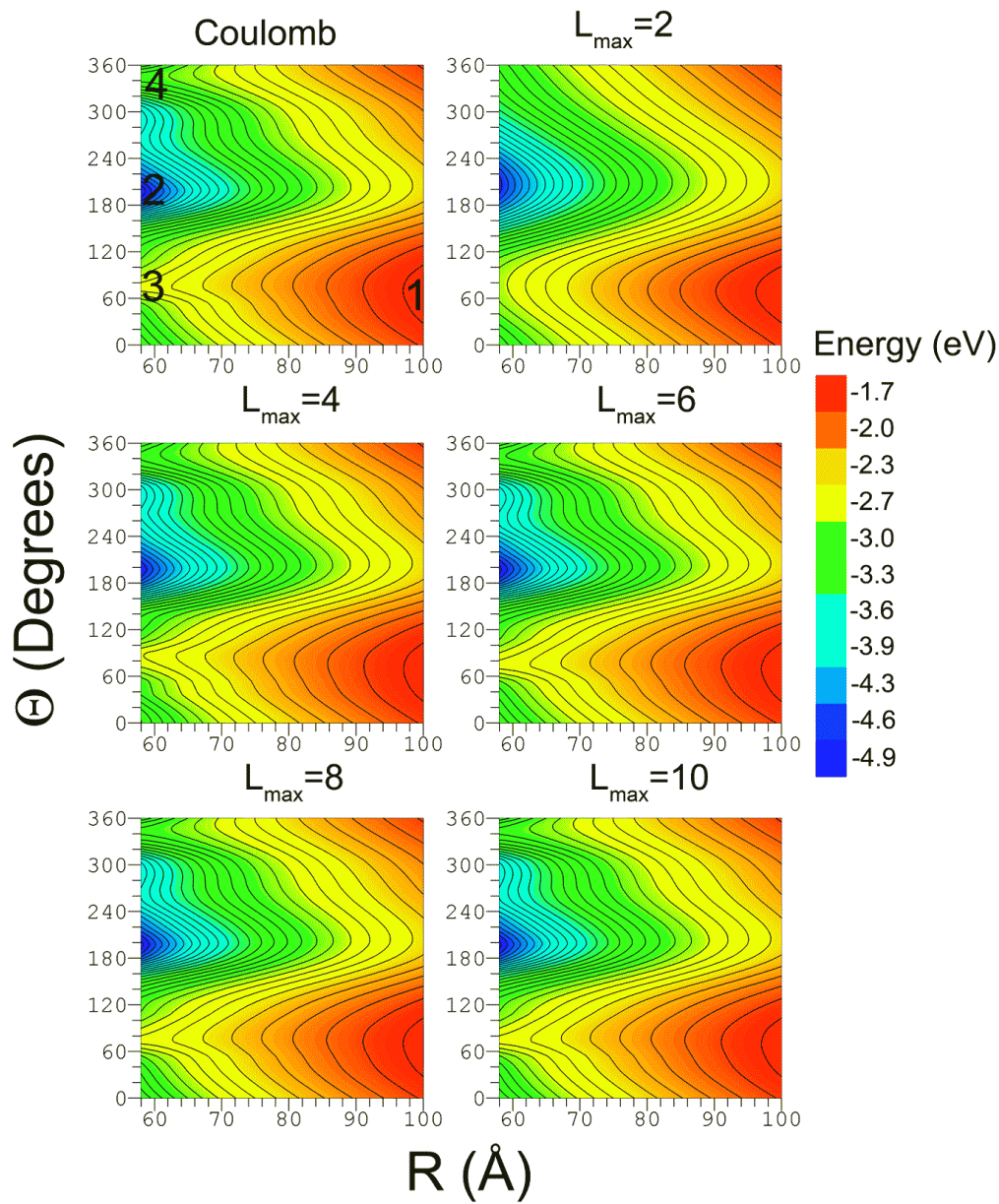}
\caption{The interaction energies of the porcine pancreatic trypsin
with the soybean trypsin inhibitor calculated as the function of
$R_{0}$ and $\Theta$ (see Fig.~\ref{belki}) within the framework of
the TCM method at different values of the principal multipole
number, $L_{max}$. The principal multipole number is given above the
corresponding image. The result of the CE calculation is shown in
the top left plot.} \label{energy_surf}
\end{figure}

The interaction energy of the porcine pancreatic trypsin with the
soybean trypsin inhibitor as the function of $R_{0}$ and $\Theta$
calculated within the framework of the TCM method is shown in
Fig.~\ref{energy_surf}. The Coulomb interaction energy between the
two proteins is shown in the top-left plot. In
\cite{SYSG06_JETP_frag} it has been shown that the interaction
energy between two well separated biological fragments arises mainly
due to the Coulomb forces. In the present paper we consider
$R_0\in[58,100]$ \AA\ and $\Theta\in[0,360]^{\circ}$, at which
condition (\ref{mainCond}) holds and both proteins can be considered
as separated. This means that the potential energy surface shown in
the top-left plot of Fig.~\ref{energy_surf} describes the
interaction energy between the porcine pancreatic trypsin and the
soybean trypsin inhibitor on the level of accuracy of 90 \% at
least.

\begin{figure}[h]
\includegraphics[scale=0.8,clip]{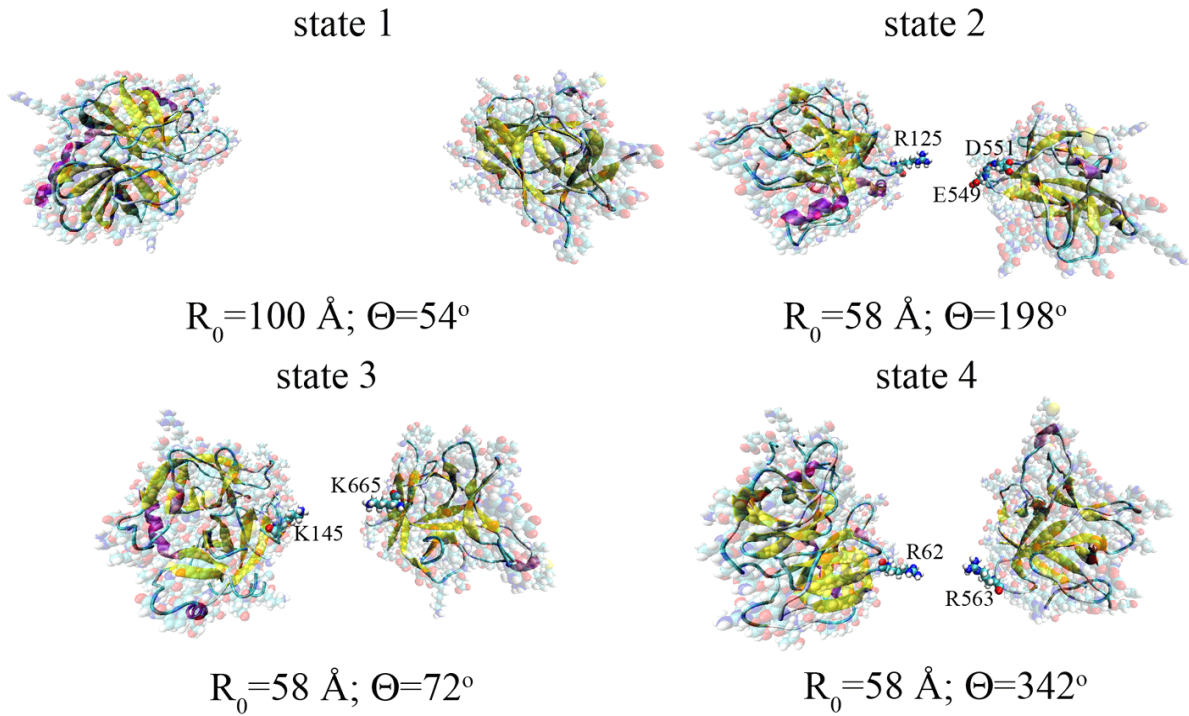}
\caption{Relative orientations of the porcine pancreatic trypsin and
the soybean trypsin inhibitor, corresponding to the selected points
on the potential energy surface presented in Fig.~\ref{energy_surf}.
Below each image we give the corresponding values of $R_0$ and
$\Theta$. Some important amino acids are marked according to their
PDB id. Figure prepared with help of the VMD visualization package
\cite{Humphrey96}} \label{fg:geometries}
\end{figure}

The top-left plot of Fig.~\ref{energy_surf} shows that one can
select several characteristic regions on the potential energy
surface marked with numbers 1-4. The corresponding configurations
(states) of the system are shown in Fig.~\ref{fg:geometries}. The
potential energy surface is determined by the Coulomb interactions
between atoms, thus at large distances it raises and asymptotically
approaches to zero. State 1 has the maximum energy within the
considered part of the potential energy surface because this state
corresponds to the largest contact separation distance between
porcine pancreatic trypsin and the soybean trypsin inhibitor being
equal to 54.8 \AA.

At smaller distances the potential energy decreases due to the
attractive forces acting  between the two proteins. State 2
corresponds to the minimum on the potential energy surface. It
arises because a positively charged polar arginine (R125) from the
porcine pancreatic trypsin approaches the negatively charged site of
the soybean trypsin inhibitor, which includes negatively charged
polar amino acids glutamic acid (E549) and aspartic acid (D551) (see
state 2 in Fig.~\ref{fg:geometries}). The strong attraction between
the amino acids leads to the formation of a potential well on the
potential energy surface. This observation is essential for dynamics
of the attachment process of two proteins, because it establishes
the most probable angle at which the proteins stick in the
(xy)-plane of the considered coordinate frame
($\Theta=192^{\circ}$).

States 3 and 4 correspond to the saddle points on the potential
energy surface and have energies higher than state 2. They are
formed because at these configurations two positively charged polar
amino acids from the two proteins become closer in space providing a
source of a local repulsive force. In state 3 these amino acids are
lysines (K145 and K665) (see state 3 in Fig.~\ref{fg:geometries}),
and in state 4 these are arginines (R62 and R563)(see state 4 in
Fig.~\ref{fg:geometries}).

In the top-right plot of Fig.~\ref{energy_surf} we present the
potential energy surface obtained within the framework of the TCM
method with $L_{max}=2$, i.e. with accounting for up to the
quadrupole-quadrupole interaction term in the multipole expansion
(\ref{Formula}). From the figure it is seen that the TCM method
describes correctly the major features of the potential energy
landscape (i.e. the position of the minimum and maximum as well as
their relative energies). However, the minor details of the
landscape, such as the saddle points 3 and 4 (see top-left plot of
Fig.~\ref{energy_surf}) are missed.

\begin{figure}[h]
\includegraphics[scale=0.82,clip]{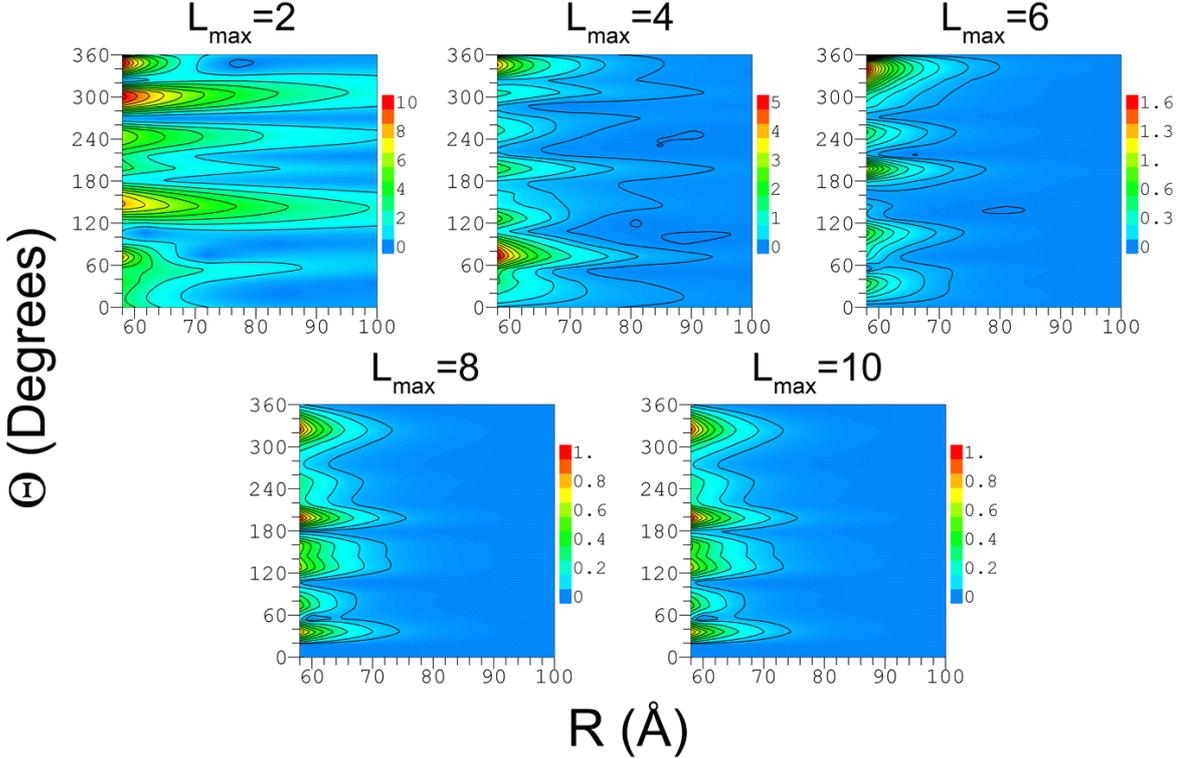}
\caption{Relative error of the interaction energies of the porcine
pancreatic trypsin with the soybean trypsin inhibitor calculated as
the function of $R_{0}$ and $\Theta$ within the framework of the TCM
method at different values of the principal multipole number,
$L_{max}$. The principal multipole number is given above the
corresponding image.} \label{fg:error}
\end{figure}

The relative error of the TCM method can be defined as follows:

\begin{equation}
\eta(L_{max})(R_0,\Theta)=\frac{|U_{coul}(R_0,\Theta)-U_{mult}^{L_{max}}(R_0,\Theta)|}{|U_{coul}(R_0,\Theta)|}\cdot100\%,
\label{eq:error}
\end{equation}

\noindent where $U_{coul}(R_0,\Theta)$ and
$U_{mult}^{L_{max}}(R_0,\Theta)$ are the Coulomb energy and the
energy calculated at given values of $R_{0}$ and $\Theta$ within the
TCM method respectively. In the top-left plot of Fig.~\ref{fg:error}
we present the relative error calculated according to
(\ref{eq:error}) for $L_{max}=2$. From this plot it is clear that
significant deviation from the exact result arise at
$\Theta\sim50-60^{\circ}$, $140-150^{\circ}$, $245^{\circ}$,
$300-310^{\circ}$ and $350^{\circ}$. The discrepancy at
$\Theta\sim50-60^{\circ}$, $\Theta\sim300-310^{\circ}$ and
$\Theta\sim350^{\circ}$ arises because the saddle points 3 and 4,
can not be described within the framework of TCM method with
$L_{max}=2$. The discrepancy at $\Theta\sim140-150^{\circ}$ and
$\Theta\sim245^{\circ}$ is due to the error in the calculation of
the slopes of minimum 2 at $R_0=58$ \AA\ and $\Theta=198^{\circ}$.

It is worth noting that the relative error of the TCM method  with
$L_{max}=2$ is less than 10 \%. With increasing distance between the
proteins, the relative error decreases, and becomes less than 5 \%
at $R_0\ge72$ \AA\ and less than 3 \% at $R_0\ge86$ \AA. This means
that already at $L_{max}=2$ the TCM method {reproduces} with a
reasonable accuracy the essential features of the potential energy
landscape. This observation is very important, because TCM method
with $L_{max}=2$ requires less computer time then the CE calculation
already at $N=4223$ (see Tab.~\ref{peresech}). Thus, the TCM method
can be used for the identification of major minima and maxima on the
potential energy surface of macromolecules and modeling dynamics of
complex molecular systems.

Accounting for higher multipoles in the multipole expansion
(\ref{Formula}) leads to a more accurate calculation of the
potential energy surface. In the second row of
Fig.~\ref{energy_surf} we present the potential energy surfaces
obtained within the framework of the TCM method with $L_{max}=4$ and
$6$. From these plots it is seen that all minor details of the
Coulomb potential energy surface, such as the saddle points 3 and 4
are reproduced correctly. Figure \ref{fg:error} shows that the TCM
method with $L_{max}=4$ gives the maximal relative error of about 5
\% at $R_0=58$ \AA\ and $\Theta=75^{\circ}$, in the vicinity of the
saddle point 3. The relative errors in the vicinity of the saddle
point 4 and minimum 2 are equal to 4 \% and 1 \% respectively. The
error becomes less then 1 \% for all values of angle $\Theta$ at
$R_0\ge70$ \AA. For $L_{max}=6$, the largest relative error is equal
to $1.5$ \% at $R_0=58$ \AA\ and $\Theta=340^{\circ}$ (saddle point
4), becoming less then 1 \% at $R_0\ge61$ \AA.

%

By accounting for the higher multipoles in the multipole expansion
(\ref{Formula}) one can increase the accuracy of the method. Thus,
with $L_{max}=8$ and $10$ it is possible to calculate the potential
energy surface with the error less then 1 \% (see bottom row in
Fig.~\ref{energy_surf} and Fig.~\ref{fg:error}). Although the time
needed for computing the potential energy surfaces with $L_{max}=8$
and $10$ is larger than the time needed for computing the Coulomb
energy directly (see Tab.~\ref{peresech}), we present these surfaces
in order to stress the convergence of the TCM method.

\section{Conclusion}
In the present paper we have proposed a new method for the
calculation of the Coulomb interaction energy between pairs of
macromolecular objects. The suggested method provides a linear
scaling of the computational costs with the size of the system and
is based on the two center multipole expansion of the potential.
Analyzing the dependence of the required computer time on the system
size, we have established the critical sizes at which our method
becomes more efficient than the direct calculation of the Coulomb
energy.

{The comparison of efficiency of the TCM method with the efficiency
of FMM allows us to conclude that the TCM method has proved to be
faster and more accurate than the classical FMM.}

The method based on the two center multipole expansion can be used
for the efficient computation of the interaction energy between
complex macromolecular systems.  To determine that we have
considered the interaction between two proteins: porcine pancreatic
trypsin and the soybean trypsin inhibitor. The accuracy of the
method has been discussed in detail. It has been shown that
accounting of only four multipoles in both proteins gives an error
in the interaction energy less than 5 \%.

{The TCM method is especially useful for studying dynamics of rigid
molecules, but it can also be adopted for studying dynamics of
flexible molecules. In this work we have developed a method for the
efficient calculation of the interaction energy between pairs of
large multi particle systems, e.g. macromolecules, being in {\it
vacuo}. The investigation of biomolecular systems consisting of
several components (i.e complexes of proteins, DNA, macromolecules
in solution) and the extension of the TCM method for these cases
deserves a separate investigation. If a system of interest consists
of several interacting molecules being placed in a solution, one can
use the TCM method to describe the interaction between the molecules
and then to take account of the solution as implicit solvent. This
can be achieved using for example the formalism of the
Poisson-Bolzmann \cite{Fogolari99,Baker01}, similar to how it was
implemented for the description of the antigen-antibody
binding/unbinding process \cite{ElsaPoster,ElsaPaper}. The other
possibility is to split the whole system into boxes and account for
the solvent explicitly by calculating the interactions between the
boxes and the molecules of interest. This can be achieved by using
the TCM method or a combination of the FMM and the TCM methods. In
this case the FMM can be used for the calculation of the resulting
effective multipole moment of the solvent, while the TCM method is
much better suitable for the description of the macromolecules
energetics and dynamics. Note that all of the suggested
methodologies provide linear scaling of the computation time on the
system size. }

The results of this work can be utilized for the description of
complex molecular systems such as viruses, DNA, protein complexes,
etc and their dynamics. Many dynamical features and phenomena of
these systems are caused by the electrostatic interaction between
their various fragments and thus the use of the two center multipole
expansion method should give a significant gain in their computation
costs.

\section{Acknowledgements}
This work is partially supported by the European Commission within
the Network of Excellence project EXCELL and by INTAS under the
grant 03-51-6170. {We are grateful to Dr. Paul Gibbon for providing
us with the FMM code. We thank Dr. Axel Arnold for his help with
compiling the programs as well as for many insightful discussions.}
We also thank Dr. Elsa Henriques and Dr. Andrey Korol for many
useful discussions. {We are grateful to Ms. Stephanie Lo for her
help in proofreading of this manuscript.} The possibility to perform
complex computer simulations at the Frankfurt Center for Scientific
Computing is also gratefully acknowledged.

\end{document}